\documentclass[%
 reprint,
 amsmath,amssymb,
 aps,
]{revtex4-1}

\usepackage{graphicx}% Include figure files
\usepackage{dcolumn}% Align table columns on decimal point
\usepackage{bm}% bold math
\usepackage{feynmp}
\usepackage[utf8]{inputenc}
\usepackage{amsmath,amssymb}
\usepackage{float}
\usepackage{amsfonts}
\usepackage{dsfont}
\usepackage{color}
\usepackage{varwidth}
\usepackage[framemethod=tikz]{mdframed}
\usepackage{epigraph}
\usepackage{tikz}

\usepackage{tikz} %For Drawing with pdflatex
\usetikzlibrary{decorations.markings} % For Arrow heads in the middle
\usetikzlibrary{decorations.text} % For text along path
\usetikzlibrary{positioning} % For relative positioning
\usetikzlibrary{backgrounds} % For background layer
\usetikzlibrary{fit} % To make background fit
\usetikzlibrary{calc} % To calculate e.g. the distance between nodes
\usetikzlibrary{shapes} % more shapes

\newcommand\be{\begin{equation}}
\newcommand\ee{\end{equation}}
\newcommand\bea{\begin{eqnarray}}
\newcommand\eea{\end{eqnarray}}

\begin{document}

%\preprint{APS/123-QED}

\title{Chiral symmetry breaking generalizes in tensor theories}

\author{P. Diaz and J. A. Rosabal}

\affiliation{Fields, Gravity and Strings @CTPU, Institute for Basic Science, 55, Expo-ro, Yuseong-gu, Daejeon, Korea, 34126.}

\date{\today}

\begin{abstract}
In this letter we uncover a new facet of chiral symmetry and the implications of its breaking in some theories. By generalizing the concept of chiral symmetry, tensor theories naturally arise. This novel approach adds to the known uses of tensor theories (quantum gravity, holography, entanglement,etc.) a  possible link to QCD phenomena. 
\end{abstract}

\maketitle

\section{\label{sec:1}Introduction}

Chiral symmetry breaking has been very successful in describing the appearance and properties of some particles in the standard model \cite{GellMann:1961ky,Nambu:1960xd}. It is known that if quarks were massless, QCD would be invariant under global transformations  $SU_R(3)\times SU_L(3)$. This group  acts on given bilinears  of the fermion fields, schematically $\Phi^{i}_j(x)=M^{\alpha}_{\beta}q^i_\alpha(x)\bar{q}^{\beta}_{j}(x)$, transforming in the fundamental-antifundamental representation of each $SU(3)$. Usually, $M=1+\gamma_5$, and the expectation value of $\Phi^{i}_j(x)$ is considered as the order parameter. 

Quarks are not massless, so chiral symmetry is manifestly broken in Nature. The pattern we observe is
 \be
SU_R(3)\times SU_L(3)\to SU_{R+L}(3),
\ee 
where the subscript $R+L$ refers to the diagonal subgroup. This pattern of symmetry breaking can be extended to $N$ flavours, with remaining diagonal group $SU(N)$ \cite{Coleman:1980mx}.

It is natural to ask whether there exists a generalized notion of chiral group and chiral symmetry breaking and which objects would hold it. 
As a motivation to generalize the chiral symmetry group, we can use the concept of polyquark boundstates \cite{Olsen:2014qna,Ali:2017jda}, which is an extension of the bilinears above-mentioned. We shall assume that there exists a hypothetical polyquark state 
\be\label{qs}
\Phi^{i_1\dots i_d}_{j_1 \dots j_d}(x)=M^{\alpha_1\dots \alpha_d}_{\beta_1\dots \beta_d}q_{\alpha_1}^{i_1}(x)\dots q_{\alpha_d}^{i_d}(x)\bar{q}^{\beta_1}_{j_1}(x)\dots \bar{q}^{\beta_d}_{j_d}(x),
\ee
where tensor  $M$ does not carry any symmetry and $i_k,j_k=1,\dots,N$, with $N>d$.
As we see, the extended notion of chiral symmetry group is associated to tensors, a connection that we explore throughout this letter.

Tensor  theories were popular in the nineties for their potential in describing quantum gravity, after the success of matrix theories describing gravity in two dimnesions \cite{DiFrancesco:1993cyw}.  
Nowadays, there is a revival of the subject \cite{Gurau:2009tw, Gurau:2010ba, Gurau:2011aq, Gurau:2011xq, Gurau:2011xp, Klebanov:2017nlk, Giombi:2017dtl, Bulycheva:2017ilt, Giombi:2018qgp, Klebanov:2018fzb, Ferrari:2017jgw, Geloun:2013kta, BenGeloun:2017vwn, Itoyama:2017emp} since they have been conjectured to be related to holography \cite{Witten:2016iux}, in the context of SYK duality \cite{ Azeyanagi:2017drg, Yoon:2017nig, Jevicki:2016bwu, Das:2017wae}. Besides, new applications of tensor theories in condensed matter physics are being developed, mainly since these theories are intimately related with the entanglement phenomenon. See, for instance, \cite{Vasseur:2018gfy} and references therein.

Spontaneous symmetry breaking (SSB) is known to generate massless bosons (Goldstone modes). In this letter we show that generalized chiral symmetry breaking in tensor theories generates Goldstone bosons which get arranged into matrices. Thus, matrix effective theories emerge naturally in the context of tensor theories with SSB. 

 Random matrices models were first used in the realm of nuclear physics \cite{wigner}. Since then, matrix models have been increasingly applied in many areas of mathematics  and physics such as number theory, string theory, quantum gravity, holography, etc. Recently, matrix models have appeared to be intriguingly related to tensor theories  \cite{Diaz:2018zbg, Bonzom:2013lda,  Krishnan:2017ztz}. However, the connection remains unclear so far.  
 
 In this letter we step forward in unravelling this relation by showing that SSB in tensor theories provides a link with matrix theories, full technical details can be found in \cite{Diaz:2018eik}. Conceptually, this connection could clarify the relation between tensor theories and quantum gravity, and holography. Besides, as an application for QCD, if tensors were built on quarks as in  \eqref{qs}, there is a chance for testing exotic bound states via the extended version of chiral symmetry breaking. These states would appear as extra octets, arranged into matrices, similar to the ordinary meson multiplet.

\section{Extended chiral group and SSB}
In this section, we define the extended chiral symmetry breaking patterns. For the sake of generality we consider unitary groups instead of $SU(N)$.
We propose the extension of the chiral symmetry $U_R(N)\times U_L(N)$ into
\be\label{gdd}
G_{d\bar{d}}(N)=\prod_{k=1}^d[U_k(N)\times U_{\bar k}(N)],
\ee
 with $d\geq 1$,  where $d=1$ reproduces the ordinary chiral symmetry. The object which naturally transforms under \eqref{gdd} is a tensor field, 
 \begin{multline}\label{tensortrans}
\Phi'^{i_1\dots i_d}_{j_1 \dots j_d}(x)\\=(g_1)_{m_1}^{i_1}\cdots (g_d)_{m_d}^{i_d} (g_{\bar{1}}^\dagger)_{j_1}^{l_1}\cdots (g_{\bar{d}}^\dagger)_{j_d}^{l_d}\Phi^{m_1 \dots m_d}_{l_1 \dots l_d}(x),
\end{multline}
 where $g_k\in U_k(N)$.

 Unlike the ordinary chiral symmetry breaking, the generalized notion contains a richer pattern structure. For instance, for $d=2$ (this could be a tetraquark state as in \eqref{qs}), we can have three different patterns
 \bea
&&\text{Diag}[U_1(N)\times U_{\bar{1}}(N)]\times \text{Diag}[U_2(N)\times U_{\bar{2}}(N)], \label{d2id} \\ 
&&\text{Diag}[U_1(N)\times U_{\bar{2}}(N)]\times \text{Diag}[U_2(N)\times U_{\bar{1}}(N)], \label{d212} \\ 
&&\text{Diag}[U_1(N)\times U_{\bar{1}}(N)\times U_2(N)\times U_{\bar{2}}(N)]\label{d2diag}, 
\eea
and the complexity increases rapidly with $d$. In general, we will have 
 \be \label{SSBpattern}
G_{d\bar{d}}(N)\longrightarrow G_\omega(N)=\prod_{\alpha=1}^\omega\text{Diag}[H_{\alpha}],
\ee
where each $H_{\alpha}$ is a tensor product of $2n_\alpha$ different unitary groups  ($n_\alpha$ barred and $n_\alpha$ unbarred) such that 
\be\label{interH1}
H_{\alpha}\cap H_\beta=\emptyset,\qquad \alpha\neq \beta,\qquad \sum_{\alpha=1}^\omega (2n_\alpha)=2d.
\ee
 Notice that $\text{Diag}[H_{\alpha}]$ is a unitary group and it will often be denoted $U_\alpha(N)$.
The symmetry breaking into the full diagonal group $G_1(N)=U(N)$ corresponds to $\omega=1$, whereas for $\omega=d$ the remaining symmetry group is $G_d(N)=\prod_{\alpha=1}^dU_\alpha(N)$. 

 For each broken continuous  symmetry there is a Goldstone mode. Thus, for the patterns \eqref{SSBpattern}, the number of Goldstone bosons that will result   is counted by subtracting the number of generators of the initial and remaining symmetry groups, i.e.,
\be\label{conteo}
d_{G_{d\bar{d}}(N)} -d_{G_{\omega}(N)} =2dN^2-\omega N^2=(2d-\omega)N^2.
\ee

\subsection{Inducing SSB}

In order to study spontaneous symmetry breaking in tensor theories we will extend the so-called $\epsilon$-term technique, developed for the scalar field in \cite{Matsumoto:1974nt, Matsumoto:1974gp}. The idea is to introduce a symmetry breaking  term ($\epsilon$-term) in the path integral. The theory is invariant under $G_{d\bar{d}}(N)$, which is stated as $\mathcal{L}\big[\Phi'(x)\big]=\mathcal{L}\big[\Phi(x)\big]$, where $\Phi'$ and $\Phi$ are related by \eqref{tensortrans}. In this letter we are not taking extra assumptions  on the Lagrangian of the theory $\mathcal{L}\big[\Phi(x)\big]$, as long as its potential allows SSB. The $\epsilon$-term, on the other hand, is chosen to be invariant only under $G_\omega(N)$.

Extending the method of \cite{Matsumoto:1974nt}, the generating functional of the tensor theory is written as
\begin{multline}\label{symPI}
Z_{\epsilon}\big[J,\overline{J}\big]=\frac{1}{\mathcal{N}}\int D\Phi D\overline{\Phi}\exp\bigg(\text{i}\int d^4x\Big\{\mathcal{L}\big[\Phi(x)\big] \\+ \overline{J}(x)\cdot \Phi(x)+J(x)\cdot \overline{\Phi}(x) +\text{i}\epsilon|\Phi(x)-v|^2  \Big \}   \bigg),
\end{multline}
 with $\mathcal{N}=Z_{\epsilon}\big[0,0\big]$, and the $\epsilon$-term is
\begin{equation}
|\Phi(x)-v|^2=\Phi(x)\cdot\overline{\Phi}(x)-\Phi(x)\cdot\overline{v}-v\cdot\overline{\Phi}(x)+v\cdot\overline{v},
\end{equation}
where the dot indicates full contraction of indices,
$X\cdot \overline{Y}=X_{i_1\dots i_d}^{j_1 \dots j_d}\overline{Y}^{i_1\dots i_d}_{j_1\dots j_d}$.

The tensor $v^{i_1\dots i_d}_{j_1 \dots j_d}$ is an invariant of $G_\omega(N)$, which determines the SSB pattern.
The $\epsilon$-term technique is intimately related to the Ward-Takahashi identities which are used to identify the Goldstone bosons of the symmetry breaking induced by $v$, \cite{Diaz:2018eik}.

 \section{SSB patterns}
 In this section we explore the structure of the $\epsilon$-term, specified by the tensor $v$, that drives the different SSB patterns \eqref{SSBpattern}. 
 In ordinary chiral symmetry breaking, the $v$ tensor which drives the breaking into the diagonal group is
 $v^i_j=v\delta^i_j$, with $v\in\mathbb{C}$, and the order parameter $\langle\Phi^i_j(x)\rangle=a\delta^i_j$,\cite{Coleman:1980mx}. For inducing the SSB patterns  \eqref{SSBpattern}, the natural extension 
is constructed using Kronecker deltas. Thus, in general, $v$ can be
\be\label{vsigma}
v^{i_1\dots i_d}_{j_1\dots j_d}=\sum_{\sigma\in S_d}v_\sigma\delta^{i_1\dots i_d}_{j_{\sigma(1)}\dots j_{\sigma(d)}},
\ee
 where
$\delta^{i_1\dots i_d}_{j_1\dots j_{d}}=\delta^{i_1}_{j_1}\cdots \delta^{i_d}_{j_d}$. We have denoted $S_d$ the group of permutations which has $d!$ elements. However, as we will argue only two terms in \eqref{vsigma} are needed in order to induce {\it any} SSB pattern of the type  \eqref{SSBpattern}. 

In order to explore the relation between $v$ and the SSB patterns, we first point out that the role of each monomial $\delta$ in  \eqref{vsigma} is to link indices in pairs, up and downstairs. Accordingly, each monomial   produces the SSB pattern
\be\label{monoassociation}
\delta^{i_1\dots i_d}_{j_{\sigma(1)}\dots j_{\sigma(d)}}\longrightarrow \prod_{k=1}^d\text{Diag}[U_k(N)\times U_{\sigma(\bar{k})}(N)],
\ee
which is $G_d(N)$ with the notation introduced in \eqref{SSBpattern}.
For $d=1$ the ordinary chiral symmetry breaking is recovered. For $d=2$, we have the patterns \eqref{d2id} or \eqref{d212}, associated to the two permutations of $S_2$. 

The sum of two or more monomials results in the intersection of the groups induced by each monomial in  \eqref{monoassociation}. For example, for $d=2$, the linear combination $v^{i_1i_2}_{j_1j_2}=v_1\delta^{i_1i_2}_{j_1j_2}+v_2\delta^{i_1i_2}_{j_2j_1}$ corresponds to the intersection of the groups \eqref{d2id} and \eqref{d212}, resulting in the pattern \eqref{d2diag}.

Because of the increasing complexity of \eqref{SSBpattern} with $d$, one might think that for  $d>2$, more monomial contributions would be needed in order to induce a given SSB pattern. Surprisingly, only two monomials are enough. 

To prove this statement we develop a diagrammatic correspondence.
In accordance with \eqref{monoassociation}, we will graphically represent the effect of the monomial in the SSB as
\be\label{simplegeneral1}
\delta^{i_1\dots i_d}_{j_{\sigma(1)}\dots j_{\sigma(d)}}\longrightarrow \begin{array}{ccc}
g_1&\dots& g_d\\
|&\dots&|\\
g_{\sigma(\bar{1})}&\dots & g_{\sigma(\bar{d})}
\end{array},
\ee   
where $g_k$ denotes a generic element of $U_k(N)$.

Following the examples of $d=2$, the two monomials are mapped as
 \be \label{simpledeltas1}
\delta^{i_1i_2}_{j_1j_2}\longrightarrow 
\begin{array}{cc}
~g_1~&~g_2~\\
~|~&~|~\\
~g_{\bar{1}}~&~g_{\bar{2}}~
\end{array}
\qquad ,
\qquad
\delta^{i_1i_2}_{j_2j_1}\longrightarrow \begin{array}{cc}
~g_1~&~g_2~\\
~|~&~|~\\
~g_{\bar{2}}~&~g_{\bar{1}}~
\end{array}.
\ee    

Diagrams representing the intersection of two groups $G_d(N)$ and $G'_{d}(N)$  will be called ``intersection diagrams.''  They are built by concatenation of two diagrams of the type  \eqref{simplegeneral1}, associated with the permutations $\sigma$ and $\sigma'$. Identical elements upstairs are joint, and the same with the barred elements downstairs. This is exemplified in Fig.\ref{simple1}, which represents the intersection of the two monomials \eqref{simpledeltas1}.
\setlength{\unitlength}{1cm}
\begin{figure}[ht]
\begin{center}
\begin{picture}(10,5)
\put(1,2){
$v_1\delta^{i_1i_2}_{j_{1}j_{2}}+v_2\delta^{i_1i_2}_{j_2j_1}~~\longrightarrow ~~\begin{array}{cccc}
        ~g_1~ & ~g_2 ~& ~g_1 ~& ~g_2 ~ \\
        ~| ~& ~| ~& ~| ~&~ | ~ \\
        ~g_{\bar{1}}~ & ~g_{\bar{2}} ~& ~g_{\bar{2}} ~& ~g_{\bar{1}}~
     \end{array}.$} 

   {\color{blue}
   \put(4.85,3.0){\line(1,0){1.55}}
  \put(4.85,3.0){\line(0,-1){.35}}
   \put(6.4,3.0){\line(0,-1){.35}}}
  
  {\color{blue}
  \put(5.5,3.15){\line(1,0){1.5}}
  \put(5.5,3.15){\line(0,-1){.5}}
   \put(7.0,3.15){\line(0,-1){.5}}}
   
    {\color{blue}
   \put(4.6,1.05){\line(1,0){2.25}}
  \put(4.6,1.05){\line(0,1){.4}}
   \put(6.85,1.05){\line(0,1){.4}}}
  
  {\color{blue}
  \put(5.3,1.15){\line(1,0){.75}}
  \put(5.3,1.15){\line(0,1){.35}}
   \put(6.05,1.15){\line(0,1){.35}}}
   
\end{picture}
\caption[]{Intersection diagram for $d=2$ with $\sigma=(1)(2)$ and $\sigma'=(12)$.} \label{simple1}
\end{center}
\end{figure}
The diagram of  Fig.\ref{simple1} contains only one loop, and will be put in correspondence with \eqref{d2diag}, that results from the intersection of \eqref{d2id} and \eqref{d212}. As we will show below, the number of loops in the diagram is $\omega$ in the remaining group $G_\omega(N)$ in \eqref{SSBpattern}. 

Let us introduce the concept of cycle structure, which completely characterizes the diagram. Cycle structure is the set of all loops, together with their lengths and the elements they involve, that fully connect the diagram. Length is the number of elements the loop involves. The key point is that, as we will prove, the cycle structure of the intersection diagrams corresponds  with the different SSB patterns. 

In order to associate an intersection diagram to \eqref{SSBpattern} we should proceed as follows. First, we draw a plain (with no numbers) diagram 
\be\label{plain}
\underbrace{
 \begin{array}{ccc|ccc}
g&\dots& g&g&\dots& g\\
|&\dots&|&|&\dots&|\\
g&\dots & g&g&\dots & g
\end{array}}_{2d \text{ slots}}.
\ee   
Second, we draw in the diagram the loops associated to a given SSB pattern, which can be read off from the set $\{H_{\alpha}\}$: for each of the $\omega$ $H_{\alpha}$ groups  we join $n_\alpha$ slots from the LHS of \eqref{plain} with $n_\alpha$ slots of the RHS of \eqref{plain} in a single loop. This turns \eqref{plain} into a plain diagram with closed loops. We would like to emphasize that such procedure always fully connect the plain diagram. This happens because $\sum_{\alpha=1}^\omega n_\alpha=d$, and the groups $H_{\alpha}$ do not intersect each other, as stated in \eqref{interH1}.

In order to complete the diagram we assign a subscript to each element according to the given SSB pattern. For this, we pick a loop, associated to some $H_{\alpha}$,  and write the labels of the different unitary groups that $H_{\alpha}$ contains. For example, for $n_\alpha=2$, we generically have 
\be 
H_{\alpha}=U_a\times U_{\bar{b}}\times U_c\times U_{\bar{d}}.
\ee
Then the labels corresponding to $H_{\alpha}$ may be chosen as in Fig.\ref{cyclelabel}.
\setlength{\unitlength}{1cm}
\begin{figure}[H]
\begin{center}
\begin{picture}(10,1.5)
$
 \begin{array}{cc|cc}
~~g~~& ~~g~~&~~g~~& ~~g~~\\
|&|&|&|\\
g& g&g& g
\end{array}\longrightarrow
\begin{array}{cc|cc}
~~g_a~~&~~ g_c~~&~~g_c~~&~~ g_a~~\\
|&|&|&|\\
g_{\bar{d}}& g_{\bar{b}}&g_{\bar{d}}& g_{\bar{b}}
\end{array}.
$  
  {\color{blue}
   \put(-7.55,1.25){\line(1,0){2.45}}
  \put(-7.55,1.25){\line(0,-1){.55}}
   \put(-5.1,1.25){\line(0,-1){.55}}}

  {\color{blue}
   \put(-6.85,1.15){\line(1,0){.85}}
  \put(-6.85,1.15){\line(0,-1){.45}}
   \put(-6,1.15){\line(0,-1){.45}}}

 {\color{blue}
   \put(-7.8,-.75){\line(1,0){1.7}}
  \put(-7.8,-.75){\line(0,1){.2}}
   \put(-6.1,-.75){\line(0,1){.2}}}

  {\color{blue}
   \put(-7.05,-.95){\line(1,0){1.65}}
  \put(-7.05,-.95){\line(0,1){.45}}
   \put(-5.4,-.95){\line(0,1){.45}}}

 {\color{blue}
   \put(-4.,1.25){\line(1,0){2.9}}
  \put(-4.,1.25){\line(0,-1){.55}}
   \put(-1.1,1.25){\line(0,-1){.55}}}

  {\color{blue}
   \put(-3.15,1.15){\line(1,0){1}}
  \put(-3.15,1.15){\line(0,-1){.45}}
   \put(-2.15,1.15){\line(0,-1){.45}}}

 {\color{blue}
   \put(-4.2,-.75){\line(1,0){1.9}}
  \put(-4.2,-.75){\line(0,1){.2}}
   \put(-2.3,-.75){\line(0,1){.2}}}

  {\color{blue}
   \put(-3.4,-.95){\line(1,0){2.0}}
  \put(-3.4,-.95){\line(0,1){.45}}
   \put(-1.4,-.95){\line(0,1){.45}}}

\end{picture}
\vspace{1cm}
\caption[]{Labelling of a piece of the plain diagram which corresponds to a loop of length 2.} \label{cyclelabel}
\end{center}
\end{figure}
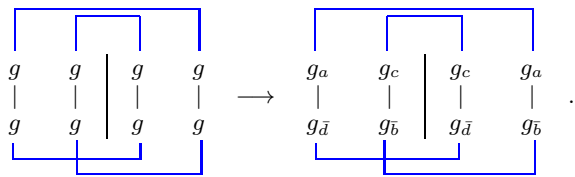
A complete diagram is obtained by applying the same prescription on each $H_{\alpha}$.
Finally, the two monomials which drive the desired SSB pattern can be read off from the full diagram. 
Now, since the SSB pattern is arbitrary, then any SSB pattern can be induced by only two monomials. This concludes the proof.

As a remark, this result suggests that a candidate for the order parameter built from the tensor field, no matter the theory, can always be taken as
\be\label{OP}
\langle\Phi^{i_1\dots i_d}_{j_1\dots j_d}(x)\rangle = a\,\delta^{i_1\dots i_d}_{j_{\sigma(1)}\dots j_{\sigma(d)}}+b\,\delta^{i_1\dots i_d}_{j_{\sigma'(1)}\dots j_{\sigma'(d)}}.
\ee

%As a remark, notice that the process of labelling is not unique. In general, there will be several pairs of deltas that would induce the same SSB pattern. This fact does not affect the conclusion of our argument since the aim is to show that any SSB pattern can be induced only with two parameters, as in  \eqref{2summands}.

 \section{Goldstone bosons $=$ matrix fields}

 In this section, we identify the massless modes that come along with the different SSB patterns and we show that they get arranged into matrix fields.
 Assuming, as usual, that there are no other massless modes   \cite{Weinberg:1996kr}, the matrix fields are the relevant content of the effective theory.
 In order to make  the link between both theories more precise, we show explicitly how Goldstone bosons are derivated from a tensor field.

 As anticipated, we make use of the powerful $\epsilon$-term technique, which we now outline, and can be found in detail in \cite{Diaz:2018eik}. This technique takes advantage of the invariance of the functional integral \eqref{symPI} under the transformations \eqref{tensortrans}. After performing \eqref{tensortrans}, \eqref{symPI} reads as a function of  $g_k=e^{\text{i}\theta_a^{k} T_a}$ and $g_{\bar{k}}=e^{\text{i}\theta_a^{\bar{k}} T_a}$ of $U_k(N)$ and $U_{\bar{k}}(N)$, respectively. This implies that the derivatives of \eqref{symPI}  with respect to the parameters $\theta_a^{k}$ and $\theta_a^{\bar{k}}$ vanish. Further functional derivatives with respect to the sources $J$ and $\overline{J}$ lead to a tower of identities among the Green functions. These are precisely the Ward-Takahashi identities.  The first functional derivatives allow to identify the Goldstone bosons \cite{Matsumoto:1974nt}.
 
 Proceeding in this way we find the Goldstone bosons $B_a^k(x)=(B^k)^i_j(x)(T_a)^i_j$. By inspection, one can check that the number of degrees of freedom of the effective theory matches \eqref{conteo}. Furthermore, they get arranged into matrices $(B^k)^i_j(x)$, which in terms of the tensor field are given by
\begin{eqnarray}\label{resultadoprincipal}
(B^{k})^i_j(x)&=&\overline{v}^{i_1\dots i_d}_{j_1\dots j\dots j_d}\Phi^{j_1 \dots i \dots j_d}_{i_1\dots i_d}(x)-v^{j_1\dots i\dots j_d}_{i_1\dots i_d}\overline{\Phi}^{i_1\dots  i_d}_{j_1\dots j \dots j_d}(x), \nonumber \\
(B^{\bar{k}})^i_j(x)&=&\Phi^{i_1\dots i_d}_{j_1\dots j\dots j_d}\overline{v}^{j_1 \dots i \dots j_d}_{i_1\dots i_d}(x)-\overline{\Phi}^{j_1\dots i\dots j_d}_{i_1\dots i_d}v^{i_1\dots  i_d}_{j_1\dots j \dots j_d}(x),\nonumber \\ 
\end{eqnarray}
where $i,j$ in the RHS of \eqref{resultadoprincipal} are in the slot $k,\bar{k}=1,\dots,d$. 

Note that \eqref{resultadoprincipal} defines $2d$ $N\times N$-matrices, which exceeds  
\eqref{conteo}. However, as explicitly shown in \cite{Diaz:2018eik}, the fields in \eqref{resultadoprincipal} are not linearly independent.  The number of linearly independent modes  match exactly \eqref{conteo}. The matrices in \eqref{resultadoprincipal} would be the generalization of the meson octet in the Standard Model if the tensor field were built from quarks as in \eqref{qs}.

 \section{Conclusion and outlook}\label{conclusions}
 In this letter, we have generalized the concept of chiral group $U_R(N)\times U_L(N)$ to $G_{d\bar{d}}(N)$  in \eqref{gdd}, and chiral symmetry breaking in \eqref{SSBpattern}. In this extended setup, the SSB patterns lead to diagonal subgroups of $G_{d\bar{d}}(N)$.  These SSB patterns are induced by $G_\omega(N)$-invariant tensors $v$, whose general form is shown in \eqref{vsigma}. In order to elucidate the intricate relation between the monomial constituents of $v$ and the SSB patterns, we develop a diagrammatic correspondence. Besides, the correspondence provides a  straightforward way of visualizing the  SSB patterns. Surprisingly, from diagram inspection, we conclude that only two (complex) parameters are needed to induce any SSB, and suggests an extremely simple form for the order parameters  \eqref{OP}. Another central result of this work is the identification of the Goldstone bosons, arranged as matrix fields. They are explicitly written in terms of the original tensor modes and $v$ in \eqref{resultadoprincipal}. 
 
We would like to highlight the possible role of the extended chiral symmetry in Nature, as it appears in different branches in physics. For instance,  at high energies, tensor theories are relevant in quantum gravity and recently, in holography via SYK. Although the relation remains still unclear. We hope that  the link to matrix models that we propose could  conceptually clarify this matter.
Nevertheless, the experimental verification of the predictions coming from tensor theories are out of reach with the current particle accelerators, whereas their effective low energy matrix theories, proposed in this letter, could in principle be observed.

A more promising scenario for testing the generalized chiral symmetry breaking is QCD. According to what we have exposed, the existence of exotic bound states in QCD could be checked.  For instance, if a tetraquark state exists as in \eqref{qs}, at low energies it should be seen as  
\begin{itemize} 
\item Either, two matrix fields  $(B^1)^i_j(x)$ and $(B^2)^i_j(x)$, transforming in the adjoint of two different diagonal $SU(3)$ groups. This would correspond to the analog of the SSB patters \eqref{d2id} or \eqref{d212}.
\item Or, three matrix fields $(B^{1,2,3})^i_j(x)$ transforming in the adjoint of the remaining diagonal group  $SU(3)$, associated to the analog to the SSB pattern \eqref{d2diag}.
\end{itemize} 
%\footnote{Only one matrix appear in this case because we are assuming that the tensor field is antisymmetric as a consequence of its fermionic constituents and the assumption that the matrix $M$ between quarks is symmetric. Usually, $M=1+\gamma_5$, which is symmetric.}
%a single   $3\times 3$-matrix field $B^i_j(x)$, similar to the ordinary octet of mesons, but with higher mass.

 Since quarks are massive, the ordinary chiral symmetry breaking is not spontaneously but manifestly broken in Nature. A similar scenario is expected for the generalized chiral symmetry. As the known mesons in QCD, the matrix fields mentioned above will not be massless. However, they could be light, what  would make them testable.

\section*{Acknowledgements}
We are grateful to Robert de Mello Koch and Junchen Rong for fruitful discussions.

\end{document}